\documentclass[draft,jgrga]{AGUTeX}

  \usepackage[dvipdf]{graphicx}

  \setkeys{Gin}{draft=false}
\authorrunninghead{YERMOLAEV ET AL.}

\titlerunninghead{GEOEFFECTIVENESS AND EFFICIENCY}

\begin{document}

%
%

\title{Geoeffectiveness and efficiency of CIR, Sheath and ICME 
in generation of magnetic storms
}
%
%

%
%




\authors{Yu. I. Yermolaev, \altaffilmark{1}
N. S. Nikolaeva , \altaffilmark{1} 
I. G. Lodkina, \altaffilmark{1} 
M. Yu. Yermolaev \altaffilmark{1}
}

\altaffiltext{1}{Space Plasma Physics Department, Space Research Institute, 
Russian Academy of Sciences, Profsoyuznaya 84/32, Moscow 117997, Russia. 
(yermol@iki.rssi.ru)}






%
%


\begin{abstract}

We investigate relative role of various types of solar wind streams in generation of magnetic storms. On the basis of the OMNI data 
of interplanetary measurements for the period of 1976-–2000 we analyze 798 geomagnetic storms with $Dst \le$ --50 nT 
and their interplanetary sources: corotating interaction regions (CIR), interplanetary CME (ICME) including magnetic clouds
 (MC) and Ejecta and compression regions Sheath before both types of ICME. For various types of solar wind we study 
following relative characteristics:  occurrence rate; mass, momentum, energy and magnetic fluxes; probability of generation 
of magnetic storm  (geoeffectiveness) and efficiency of process of this generation. 
Obtained results show that despite magnetic clouds have lower occurrence rate and lower efficiency 
than CIR and Sheath they play an essential role in generation of magnetic storms due to 
higher geoeffectiveness of storm generation (i.e higher probability to contain large and long-term southward IMF $Bz$ component).

\end{abstract}

%
%

%

\begin{article}

%
%

\section{Introduction}

One of key issues of the solar-terrestrial physics is investigation of mechanisms of energy transfer from the solar wind 
into the magnetosphere and of excitation of magnetospheric disturbances.  As has been discovered by direct space 
experiments in the beginning of 1970s, the basic parameter leading to magnetospheric disturbances is negative 
(southward) Bz component of interplanetary magnetic field (IMF) (or electric field $Ey = Vx \times Bz$) 
\citep{Dungey1961,FairfieldCahill1966,RostokerFalthammar1967,Russelletal1974,Burtonetal1975,Akasofu1981}.
Numerous investigations demonstrated that IMF in the undisturbed solar wind lies in the
ecliptic plane (i.e., Bz  is close to zero) and only disturbed types of the solar wind streams can have a considerable 
value of IMF Bz. The interplanetary CME (ICME) with a compression region Sheath before 
it and the compression region 
between slow and fast solar wind streams (Corotating Interaction Region, CIR) belong to such types of  solar wind streams 
(see reviews and recent papers, for instance, by 
\cite{Tsurutanietal1988,TsurutaniGonzalez1997,Gonzalezetal1999,YermolaevYermolaev2002,HuttunenKoskinen2004,EcherGonzalez2004,YermolaevYermolaev2006,BorovskyDenton2006,Dentonetal2006,Huttunenetal2006,Yermolaevetal2007a,Yermolaevetal2007b,Yermolaevetal2007c,Pulkkinenetal2007a,Pulkkinenetal2007b,Zhangetal2007,Turneretal2009,Yermolaevetal2010a,Yermolaevetal2010b,Yermolaevetal2010c,Yermolaevetal2010d,Yermolaevetal2011,Nikolaevaetal2011,Alvesetal2011,Echeretal2011,Gonzalezetal2011,Guoetal2011,MustajabBadruddin2011} and references therein).

Experimental results showed that the magnetospheric activity induced by different types of interplanetary streams is different 
\citep{BorovskyDenton2006,Dentonetal2006,Huttunenetal2006,Pulkkinenetal2007a,PlotnikovBarkova2007,Longdenetal2008,Turneretal2009,Despiraketal2009,Despiraketal2011,Guoetal2011}.
This fact indicates that it is necessary to take into account the influence of other (in addition to IMF $Bz$    and electric field $Ey$) parameters of the solar wind, dynamics of parameter variation, and different mechanisms of generating the magnetospheric disturbances at different types of solar wind streams.
Several recent papers analyzed separately CIR, Sheath and body of ICME  and compared them with each other 
\citep{HuttunenKoskinen2004,YermolaevYermolaev2006,Huttunenetal2006,Yermolaevetal2007a,Yermolaevetal2007b,Yermolaevetal2007c,Pulkkinenetal2007a,YermolaevYermolaev2010,Yermolaevetal2010a,Yermolaevetal2010b,Yermolaevetal2011,Alvesetal2011,Despiraketal2011,Nikolaevaetal2011,Guoetal2011}.

Papers mentioned above are devoted studying of response of magnetosphere to interplanetary drives and use word {\it geoeffectiveness} 
for a designation of this link.   
It should be noted that there is a double meaning of
the term {\it geoeffectiveness}. In one case, {\it geoeffectiveness}
implies a probability with which selected phenomenon can cause a magnetic storm, i.e., the ratio between the number 
of events of a chosen type resulting in a magnetic storm and the total number of these events. 
In the other case, {\it geoeffectiveness} implies the
efficiency of storm generation by unambiguously
interrelated phenomena, i.e., the ratio between the
''output'' and ''input'' of a physical process, for example, between the values of the $Dst$ index and the  
southward  IMF Bz component. 
Below we will use the term {\it geoeffectiveness} for a designation of probability of relation between the phenomena and 
the term {\it efficiency} for a designation of efficiency of process relating phenomena.

A considerable quantity of papers is devoted investigations of geoeffectiveness
in generation of magnetic storm. The great bulk of works studies 
geoeffectiveness of magnetic clouds, and geoeffectiveness of other phenomena is studied rather poorly (see, for example, one of recent reviews by 
\cite{YermolaevYermolaev2006,YermolaevYermolaev2010,Alvesetal2011}. 
So, one of the main aims of current paper is 
investigation of geoeffectiveness of various interplanetary drivers and comparison of them to each other. 

Efficiencies of various interplanetary drivers vary with the
type of solar wind streams and may be estimated as the ratio of measured
energy output to estimated energy input (see, for example,
paper by 
\cite{Turneretal2009,Yermolaevetal2010c} 
and references therein).
In our investigations, we use Bz (Ey) and magnetospheric indices
$Dst$, $Dst^*$ (pressure corrected $Dst$), $Kp$ and $AE$ as ''input''
and ''output'' of the storm generation processes for the
estimation of efficiency of interplanetary drivers. 

\section{Methods} 

When the types of solar wind streams were classified, we
used OMNI database (see http://omniweb.gsfc.
nasa.gov \citep{KingPapitashvili2004}) for interval 1976-2000 and available
world experience in identification of solar wind streams
and the standard criteria for following parameters: velocity $V$,
density $N$, proton temperature $T$, ratio of thermal to
magnetic pressure ($\beta$-parameter), ratio of measured temperature
to temperature calculated on basis of average
"velocity–-temperature" relation $T/Texp$ \citep{Lopez1987}, thermal pressure
and magnetic field. This method allows us to identify
reliably 3 types of quasi-stationary streams of the solar
wind (heliospheric current sheet (HCS), fast streams from the
coronal holes, and slow streams from the coronal streamers),
and 5 disturbed types (compression regions before 
fast streams (CIR), and interplanetary manifestations
of coronal mass ejections (ICME) that can include
magnetic clouds (MC) and Ejecta with the compression
region Sheath preceding them). In contract
with Ejecta, MCs have lower temperature, lower ratio
of thermal to magnetic pressure ($\beta$-parameter) and
higher, smooth and rotating magnetic field 
\citep{Burlaga1991}. 
In addition,
we have included into our catalog such events (rare
enough) as direct and reverse shocks, and the rarefaction
region (region with low density) Rare but these types of events are not analyzed in this paper. 

In order to calculate yearly averaged values, we have
taken into consideration that the OMNI database contains
gaps of the data from 0 to 50\% time of year. This
procedure has been made in the assumption that  
occurrence  rate of given type of solar wind streams is similar both in intervals
of data presence and in intervals of data gap. If
during chosen year the number of events of selected solar wind 
type $Ne$ has been registered in interval of data presence $t_d$
the normalized number of the given solar wind type $Ne^*$ in this year
was defined by multiplication of occurrence rate of the
given solar wind type $Ne/t_d$ to total duration of year $t_y$, i.e. $Ne^* =  (Ne/t_d)*t_y$. 
Normalized number of solar wind events is used only for studying the time variations in occurrence rate 
of various types of streams and measured number of events is used to calculate geoeffectiveness of types of events.  
When we analyzed
durations of different types of solar wind streams, we selected intervals
of types of streams which have not data gaps at both edges
of the intervals.

Definite types of the solar wind streams were put in correspondence to all magnetic storms for which
measurements of the parameters of plasma and magnetic field in the interplanetary medium were available. 
This was done using the following algorithm. If the moment of minimum in the $Dst$ index from the list
of magnetic storms falls within the time interval of a solar wind event or is apart from it 
by no more than 2 h interval, the corresponding solar wind type is
ascribed to this storm. It should be noted that, according to
the results of analysis of 64 intense ($Dst <$ --85 nT) magnetic storms in the period 1997--2002, the average
time delay between $Dst$ peak and southward IMF Bz component is equal to  $\sim 2$ h  
\citep{GonzalezEcher2005}. 
Similar results were obtained in papers by 
\cite{Yermolaevetal2007a,Yermolaevetal2007c}. 
Thus, two hours correspond to the average time delay between
the $Dst$ peak of an intense magnetic storm and the
associated peak in the southward IMF Bz component.

In order to investigate the
dynamic relation between development of parameters
in interplanetary sources and in the magnetospheric
indices we apply the method of double
superposed epoch analysis (DSEA) \citep{Yermolaevetal2010c,Yermolaevetal2010d}. Two reference
times are used in this method: we put together the time
of storm onset (time ''0'') and
time of $Dst$ index minimum (time ''6''), the data
between them we compress or expand in such a way
that durations of the main phases of all magnetic
storms is equal to each other. This DSEA method allows us to simultaneously
study interplanetary conditions resulting in the
beginning and end of magnetic storms as well as dynamics
(temporal variations) of parameters during main phase for storms with different
durations.

\section{Results}

Obtained results are presented in 3 subsections devoted to (
1) observational statistics of  various types of solar wind streams, 
(2) probability of magnetic storm generation by these interplanetary drivers, and 
(3) efficiency of magnetic storm generation by various drivers. 

\subsection{Occurrence rate of different types of solar wind streams}

In order to estimate geoeffectiveness of different types of solar wind streams it is necessary to have a total list  
of these types of streams during sufficiently large time interval and with sufficiently large statistics. 
Measured and normalized numbers per year, average durations, temporal
parts in total times of observations as well as average values and their standard deviations of several
plasma and magnetic field parameters for various solar wind types have been presented in our publication
\citep{Yermolaevetal2009,Yermolaevetal2010a,Yermolaevetal2010b,Yermolaevetal2010c,Yermolaevetal2010d,Yermolaevetal2011}. 
It should be noted that both types of
compressed regions (CIR and Sheath) have very close
values of parameters while the parameters for 2 types of
ICME (Ejecta and MC) are different. 
In figure 1 we present yearly average values of sunspots (top panel) and yearly average distributions of times of observations 
for different types of solar wind streams (bottom panel). 
Data for different types of streams are showed by various color columns (see designation on the right of the figure) 
with height proportional to percent of observation time. 
On the average the quasi-steady types of solar wind streams 
(Fast, Slow and HCS) contain about 60\% of all solar wind observations near the Earth but time of disturbed types 
of streams decreases down to 25\% during  solar minimum and increases up to 50\% during solar maximum. 

Various types of solar wind  streams transport different values of mass, momentum, energy 
and magnetic field 
from the Sun to the Earth.  
Figure 2 shows average distributions (percentage) of values (red columns) and total Sun$'$s losses 
(parameters integrated over time, blue columns) 
mass, momentum, energy and magnetic fluxes for different types of solar wind streams.  High average values for mass, momentum, and energy fluxes are observed in compressed regions CIR and Sheath and magnetic flux in MC, but their total losses are higher in steady types of streams (Fast and Slow) than in disturbed types of streams.
In following sections of the paper we will analyze how occurrence rate of different types of streams and mass, momentum, 
energy and magnetic field transferred by these streams influence generation of magnetic storms.

\subsection{Geoeffectiveness of interplanetary drivers} 

For the entire period of time 1976--2000, 798 moderate and strong magnetic storms
with the intensity $Dst \le$ --50 nT were observed on the Earth 
 (see figure 3). But only for 464 magnetic
storms (i.e., for 58\% of all magnetic storms) corresponding various events were found in the solar wind. The sources of 
other 334 magnetic storms (i.e., of 42\% of 798 storms, grey columns in figure 3) are undeterminate, and this fact is mainly 
connected with the lack of data on plasma and interplanetary magnetic field which makes impossible to identify the solar 
wind type for magnetic storm intervals. Figure 4 presents distribution of storms when we excluded IND storms from analyses. 

Analyses of data in Figures 1 and 4 allows us to compare number of each type of solar wind streams 
and number of magnetic storms induced by these types of streams and to calculate probability 
(geoeffectiveness) of generation of magnetic storms by each types of these interplanetary drivers 
(see Table 2). Though the statistics of annual numbers of solar wind streams is small, 
the available data speak in favour of suggestion that geoeffectiveness does not change essentially 
during solar cycle. 

\subsection{Efficiency of interplanetary drivers} 

One of important problems of connection between interplanetary
conditions and magnetospheric processes is the dependence
of magnetospheric activity on temporal evolution
of solar wind plasma and IMF parameters including Bz and Ey. We found a
consistency between time evolution of cause (Bz and Ey) and
time evolution of effect ($Dst$, $Dst^*$ (pressure corrected Dst), $Kp$ and $AE$ indices) for
the time interval of ''0''--''6'' as dependence of indices on integral value of sources, for
example, 
$Dst^i.vs.Ey(\sum)^{i}= \int_{0}^{t^{i}} Ey(\tau) d\tau = \sum_0^i Ey^{k}, i=0,...,6; k=0,...,i$.

Dependencies of $Dst$ (or $Dst^*$) on the
integral of Bz (or Ey) over time are almost linear and parallel
for different types of drivers. This fact can be considered
as an indication that time evolution of main phase of storms
depends not only on current values of Bz and Ey, but also
on their prehistory. The differences between these lines are
relatively small ($\mid \Delta Dst \mid <$ 20 nT). Nevertheless we can make
following comparisons. 
For various drivers we approximated data near central parts dependencies by linear functions and 
calculated values of $Dst$ (or $Dst^*$) at fixed values of integral of $Bz$ and integral of $Ey$ 
($\int_{0}^{t} Bz(\tau) d\tau = -30$ h*nT  and 
$\int_{0}^{t} Ey(\tau) d\tau = 12$ h*mV/m) (see Table 3). 
It should be noted that used value of integral of $Ey$ is located near threshold of generation of 
magnetic storms with $Dst \le -50$ nT (i.e. used interval of integral of $Ey$ contains data 
for almost all magnetic storms)
\citep{Nikolaevaetal2012}.
Taking into account that difference in ''efficiency coefficients'' for various drivers are 
mathematically significant when they differ more 
than 10\% (i.e. 0.25 nT/(h*nT) for $Bz$ and 0.5 nT/(h*mV/m) for $Ey$), 
it is possible to note that:  
(1)	Dependencies of $Dst$ (or $Dst^*$) on the integral of Bz (or Ey) are higher in CIR, Sheath and Ejecta, than in MC, 
(i.e., efficiency of MC for process of magnetic storm generation is the lowest one);
(2)	Efficiency of  CIR, Sheath and Ejecta are closed to each other. 
Dependencies of $Kp$ (and $AE$) on
integral of Bz (and Ey)  are nonlinear (there is the
saturation effect for $AE$ index) and nonparallel. Nevertheless we made the same procedure for them as for $Dst$ and 
 $Dst^*$ indices and calculated estimations of efficiency for different drivers. Efficiency for $Kp$ and $AE$ indices is higher 
for CIR and Sheath than for MC and Ejecta.   

\section{Discussion and conclusions} 

The quantity of Sun$'$s energy flowing in a magnetosphere and causing magnetospheric
 disturbances, is defined by following processes and relations: 

1.	relative occurrence rate of disturbed types of solar wind streams (interplanetary drivers of magnetic storms), 

2.	typical values of plasma and field parameters in these types of streams,

3.	probability in magnetic storm generation  (geoeffectiveness) for these drivers (i.e. probability of occurrence of southward IMF Bz component in these drivers), and 

4.	efficiency of physical process of magnetic storm generation for various drives.

On the basis of OMNI data during 1976--2000 we estimated and compared these processes and relations. 

The results of our identification of solar wind streams were partially compared
with tabulated data of various events presented on
the websites http://star.mpae.gwdg.de/, http://lempfi.gsfc.nasa.gov/, and
with the ISTP Solar Wind Catalog on the website
http://www-spof.gsfc.nasa.gov/scripts/sw-cat/Catalog-
events.html. and presented in papers by 
\cite{CaneRichardson2003,Alvesetal2006,KoskinenHuttunen2006,Echeretal2006,Zhangetal2007}.
This comparison showed a good agreement in more than
90\% of events. 
It is important to note that, unlike numerous
papers where solar wind identifications were made
for selection of only one or two stream types we realized this approach
with a single set of criteria to eight large-scale stream
types. The obtained statistical characteristics and distributions of
the solar wind and IMF parameters in various types of
the streams well agree with previously obtained results

During the full time from 1976 to 2000 the different types of the solar wind were observed: HCS for $6 \pm 4$\%, 
MC for $2 \pm 1$\%, Ejecta for $20 \pm 6$\%, Sheath before Ejecta for $8 \pm 4$\%, Sheath before MC for $0.8 \pm 0.7$\%, 
and CIR for $10 \pm 3$\% of the total observation time.
About 53\% of the entire observation time fell on fast
 and slow solar wind (21.5\% and 31.5\% of time, respectively) (see Figure 1 and Table 1).
Large values of mass, momentum and energy are transported from the Sun to the Earth by CIR and Sheath, and of magnetic 
field by MC (see Figure 2).  

Probabilities that conditions in the interplanetary space allow to input solar wind energy to 
magnetosphere and generate magnetic storm with $Dst\le$ --50 nT are about 55\% for MC (63\% for MC with Sheath), 
about 20\% for CIR, about 8\% for Ejecta (21\% for Ejecta with Sheath) and 15\% for Sheath (see Table 2). 
Because of different occurrence rates of different solar wind streams it was found that 35\% storms were generated by Ejecta with/without Sheath, 31\% by CIR and 24\% by MC with/without Sheath (about 20\% by Sheath before MC and Ejecta). 
Taking into account dependence of numerical estimation on used method of data analysis , values of geoeffectiveness  obtained by us for MC and Ejecta (both with Sheath and without Sheath) are in a good  agreement with previous result (see review by 
\cite{YermolaevYermolaev2010}). 
Our estimation of CIR geoeffectiveness (about 20\%) is lower than one
 obtained early by 
\cite{Alvesetal2006}. 
 
Our estimations show that efficiency of MC for process of magnetic storm generation (for $Dst$ and $Dst^*$ indices) is the 
lowest one and efficiency for $Kp$ and $AE$ indices is higher for CIR and Sheath than for MC and Ejecta. 
Higher efficiency of magnetic storms generation process by Sheath than MC are discussed in several papers 
\citep{HuttunenKoskinen2004,Huttunenetal2006,Yermolaevetal2007a,Yermolaevetal2007b,Yermolaevetal2007c,Yermolaevetal2010c,Pulkkinenetal2007a,Turneretal2009,Guoetal2011}.    
Our results confirm this conclusion.   

Thus obtained results show that despite low occurrence rate and low efficiency of magnetic clouds 
they play an essential role in generation of magnetic storms due to high geoeffectiveness of storm  
generation (i.e high probability to contain large and long-term southward IMF $Bz$ component). 
Geoeffectivenesses of CIR and Sheath are lower but they are compensated by higher occurrence rate 
and efficiency.


%
%
%
%
%
%
%

\begin{acknowledgments}
The authors are grateful for the possibility of using the OMNI database. The OMNI data were obtained from the GSFC/SPDF OMNIWeb on the site http://omniweb.gsfc.nasa.gov. This work was supported by the Russian Foundation for Basic Research, projects nos. 07--02--00042 and 10--02--00277a, and by the Program 16 of Physics Department of Russian Academy of Sciences (OFN RAN).
\end{acknowledgments}

\end{article}


%
%

%
%
%
%
%


 \begin{figure}
 \noindent\includegraphics[width=12cm]{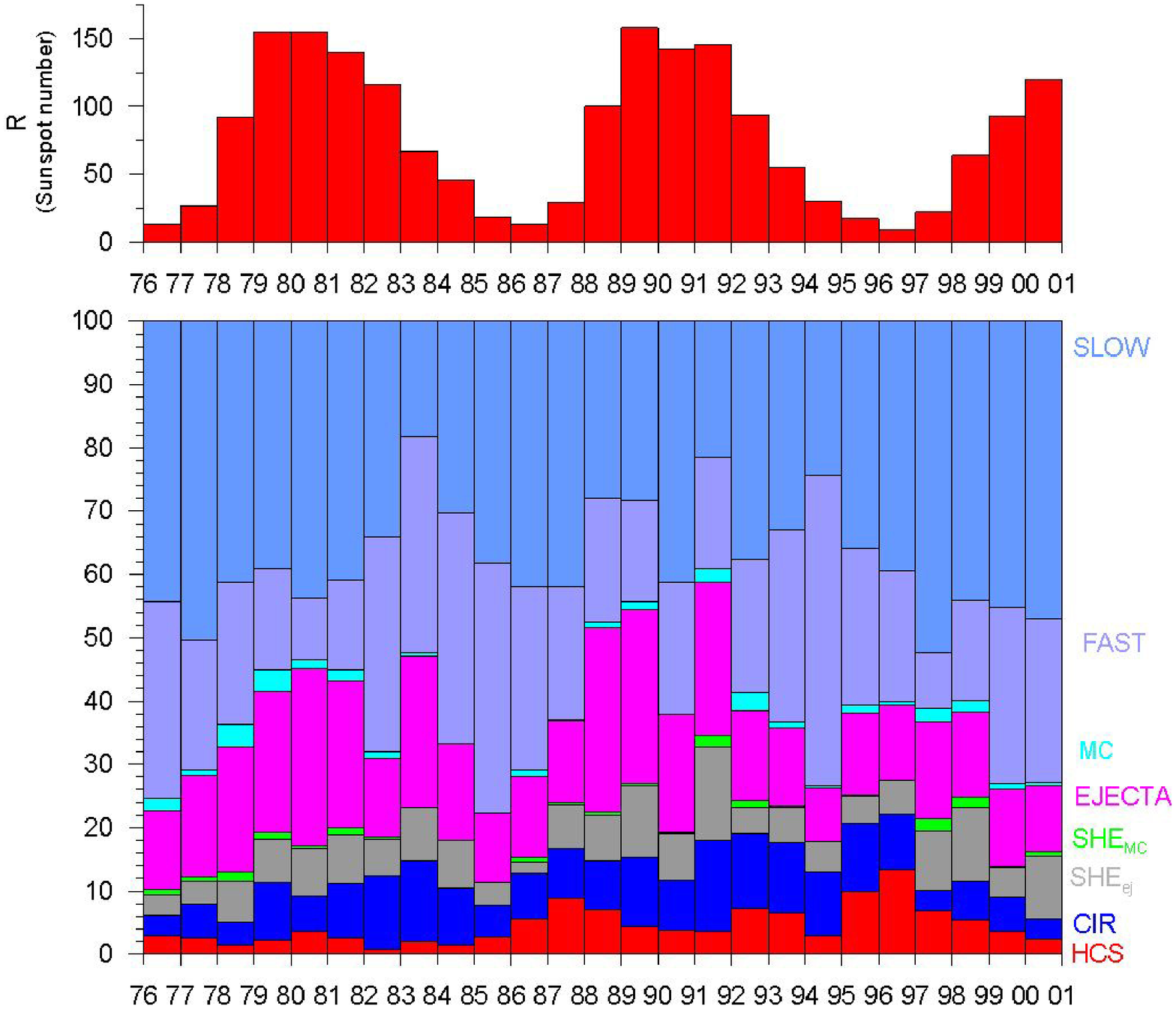}
 \caption{Yearly average values of sunspots (top panel) and yearly average distributions of times of observations for different types of solar wind (\%, bottom panel).}
 \end{figure}


 \begin{figure}
 \noindent\includegraphics[width=15cm]{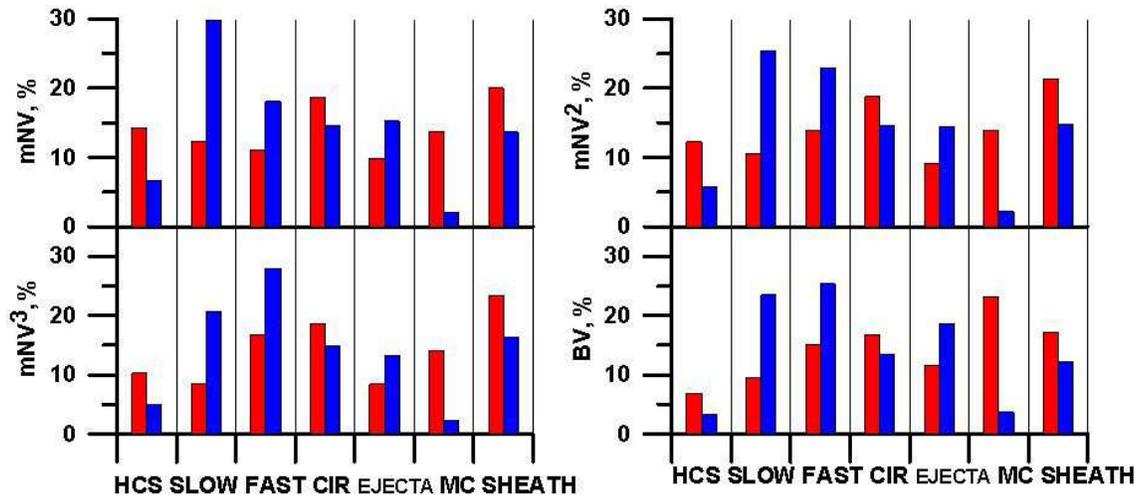}
 \caption{Average values (red columns) and Sun$'$s losses (blue columns) mass, momentum, energy and magnetic fluxes for different types of solar wind streams (\%).  }
 \end{figure}

 \begin{figure}
 \noindent\includegraphics[width=12cm]{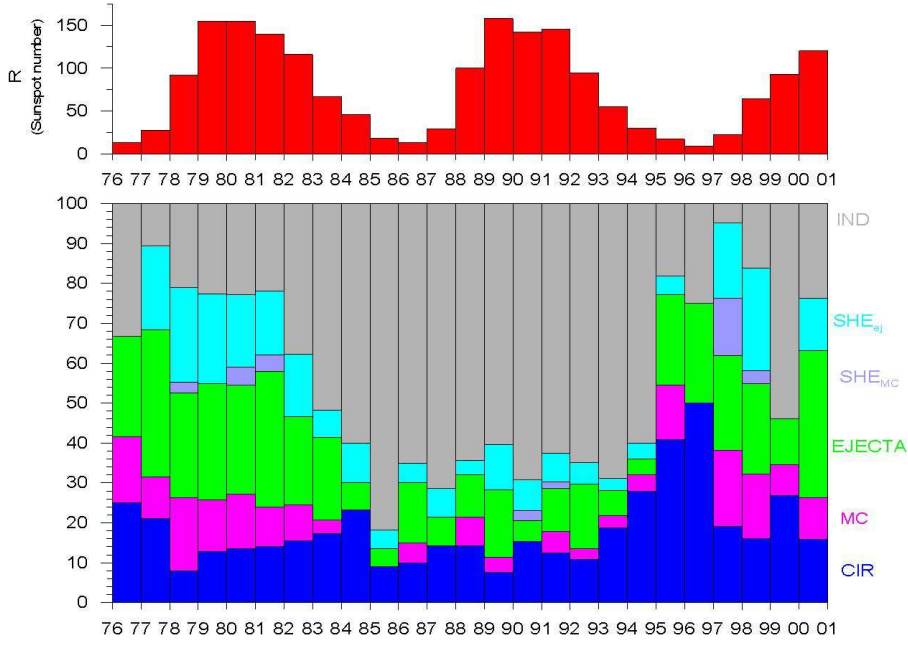}
 \caption{Sunspot number (top panel) and year-averaged distributions of magnetic storms with $Dst <$ --50 nT over types of their interplanetary drivers (\%, bottom panel). }
 \end{figure}

 \begin{figure}
 \noindent\includegraphics[width=12cm]{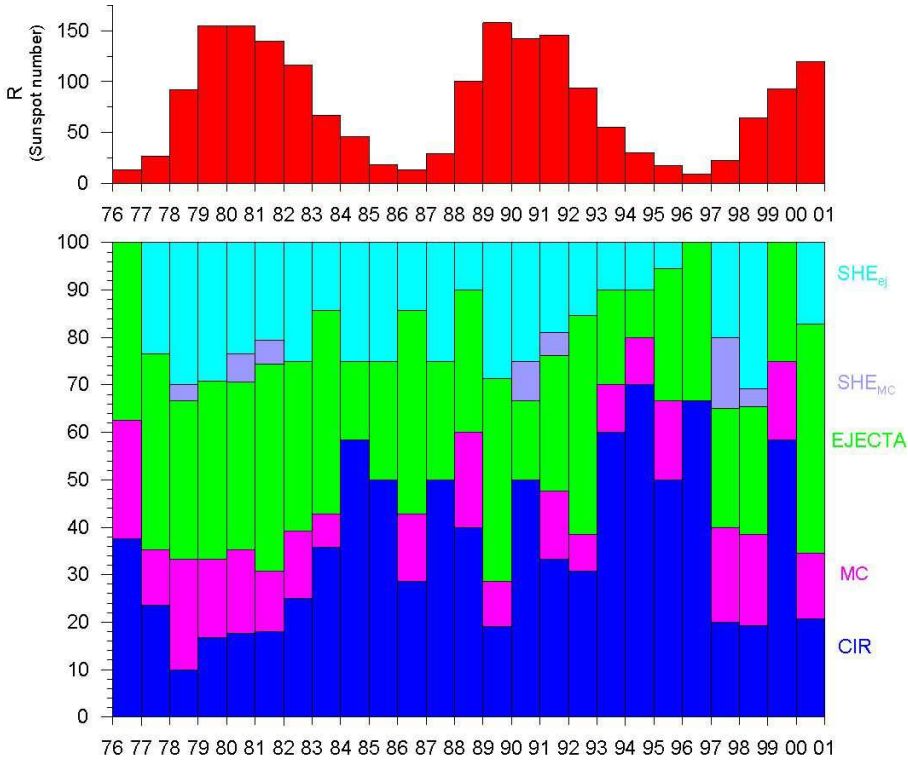}
 \caption{The same as in Figure 3 when IND storms was excluded from analyses}
 \end{figure}

%

\begin{table}
\caption{Time observation of different types of solar wind streams during 1976--2000 
}
\centering
\begin{tabular}{l c}
\hline
 Types of solar wind  & Time observations, \%  \\
\hline
  Slow                    & 31 $\pm$ 7   \\
  Fast                    & 21 $\pm$ 8   \\
  HCS                     & 6 $\pm$  4   \\
  CIR                     & 10 $\pm$ 3   \\
  Ejecta                  & 20 $\pm$ 6   \\
  MC                      & 2 $\pm$  1   \\
  Sheath before Ejecta    & 8 $\pm$  4    \\
  Sheath before MC        & 0.8 $\pm$ 0.7 \\
\hline
\end{tabular}
\end{table}

\begin{table}
\caption{Probability of generation of magnetic storms with $Dst \le -50 nT$ (geoeffectiveness) for different types of solar wind streams during 1976--2000 
}
\centering
\begin{tabular}{l cccc}
\hline
 Types of solar wind  & Number of observations & Number of storms & Part from  & Geoeffectiveness  \\
                      & of interplanetary   &  induced by this & identified &                   \\
                      & events              &  type of events  &  storms, \%&                   \\
\hline
  CIR                     & 717             & 145              & 31.2       & 0.202   \\
  Sheath before MC        & 79              & 12               & 2.6        & 0.142   \\
  Sheath before Ejecta    & 543             & 84               & 18.1       & 0.155    \\
  MC with Sheath          & 79              & 50               & 13.4       & 0.633    \\
  MC without Sheath       & 22              & 12               & 2.6        & 0.545    \\
  Ejecta with Sheath      & 543             & 115              & 24.8       & 0.212    \\
  Ejecta without Sheath   & 585             & 46               & 9.9        & 0.078    \\
\hline
\end{tabular}
\end{table}

\begin{table}[b]
\caption{Ratio of magnetospheric indices to integrated IMF $Bz$ and $Ey$ fields \\
(at fixed values of  
$\int_{0}^{t} Bz(\tau) d\tau = -30$ h*nT  and 
$\int_{0}^{t} Ey(\tau) d\tau = 12$ h*mV/m) \tablenotemark{a}
}
\centering
\begin{tabular}{l| cccc|cccc}
\hline
 SW type & $Dst/Bz$ & $Dst^*/Bz$& $Kp/Bz$& $AE/Bz$& $Dst/Ey$& $Dst^*/Ey$& $Kp/Ey$& $AE/Ey$ \\
\hline
 $CIR$      & 2.4 & 2.8& 0.18& 22.7& 5.0& 6.8& 0.45& 56.8 \\ 
 $Ejecta$   & 2.6 & 2.6& 0.17& 22.0& 6.1& 6.8& 0.43& 53.8 \\
 $MC$       & 1.9 & 2.1& 0.17& 22.3& 4.3& 4.9& 0.42& 54.2 \\
 $Ejecta+MC$& 2.3 & 2.6& 0.17& 21.8& 5.3& 6.0& 0.42& 53.3 \\
 $Sheath$   & 2.4 & 3.0& 0.20& 24.3& 4.9& 6.3& 0.46& 57.9 \\
 $IND$      & 2.9 & 2.6& 0.18& 24.0& 6.5& 6.1& 0.44& 48.9 \\
\hline
\end{tabular}
\tablenotetext{a}{Dimensions of coefficients: [$Dst/Bz$, $Dst^*/Bz$, $AE/Bz$] = nT/(h*nT), 
[$Kp/Bz$] = 1/(h*nT), 
[$Dst/Ey$, $Dst^*/Ey$, $AE/Ey$] = nT/(h*mV/m), and [$Kp/Ey$] =  1/(h*mV/m) }
\end{table}

\end{document}